\documentclass[preprint,proceedings]{rmaa}

\suppressfulladdresses 

\usepackage{paralist}

\usepackage{psfrag,color}
\usepackage{html}

\newcommand{\chandra}{{\it CHANDRA}}

\newcommand{\asca}{{\it ASCA}}

\newcommand{\xmm}{{\it XMM}}
\newcommand{\sax}{{\it BeppoSAX}}
\newcommand{\ec}{$\eta$~Carinae}

\SetYear{2006}
\SetConfTitle{Massive Stars: Fundamental Parameters and Circumstellar Interactions}

\title{Eta Carinae and other Luminous Blue Variables} 

\author{
  M. F. Corcoran \altaffilmark{1,2}}

\altaffiltext{1}{CRESST and X-ray Astrophysics Laboratory NASA/GSFC, Greenbelt, MD 20771, USA. (corcoran@milkyway.gsfc.nasa.gov).}

\altaffiltext{2}{Universities Space Research Association, 10211 Wincopin Circle, Suite 500 Columbia, MD 21044, USA.}

\shortauthor{Corcoran}
\shorttitle{Eta Car and other LBVs}

\listofauthors{Corcoran}
\indexauthor{Corcoran, M. F.}

\abstract{Luminous Blue Variables (LBVs) are believed to be evolved, extremely massive stars close to the Eddington Limit and hence prone to bouts of large-scale, unstable mass loss. I discuss current understanding of the evolutionary state of these objects, the role duplicity may play and known physical characteristics of these stars using the X-ray luminous LBVs Eta Carinae and HD 5980 as test cases. }

\addkeyword{Stars: variables}
\addkeyword{Stars: early-type}
\addkeyword{Stars: evolution}
\addkeyword{Stars: Binaries}
\addkeyword{Stars: Mass loss}
\addkeyword{X-rays: stars}
\addkeyword{Stars: Eta Carinae}

\begin{document}
\maketitle

\section{Introduction}
\label{sec:intro}

The evolution of massive stars is one of the most complex problems in modern astrophysics. The old, simple idea of a core nuclear furnace merrily burning its way down to the iron catastrophe surrounded by a relatively inert, non-magnetic envelope blissfully unaware of this impending calamity has morphed into a combined problem of core-envelope evolution intrinsically coupled through  exchange and loss of angular momentum, a process which is itself largely dependent on the as-yet poorly understood magnetic field threading the stellar interior.  That massive stars possess magnetic fields is no longer a matter of much controversy. If these fields are not simply left over and intensified from the protostellar collapse, then the resilient astronomer has numerous means at hand to create them.  The observational detection of such fields is of course a classic, difficult problem.  A breakthrough has been the recognition of a certain class of hot stars (like $\theta^{1}$ Ori  C, HD 191612, and $\tau$ Sco) which show variable spectropolarimetric or hard X-ray signatures for which there are few good alternative explanations but magnetic fields. 

And we haven't even mentioned the most fundamental problem of all: mass loss.  Stellar winds certainly drive off the lion's share of material prior to the supernova explosion in all but the most massive of massive stars ($10<M_{\mbox{main sequence}}/M_{\odot}<30$).  Above this, instabilities produced as the star evolves towards two important limits (Eddington and Humphreys-Davidson) produce in some as-yet unspecified way giant outbursts of material (perhaps removing as much as 50-90\% of the outer layers of the star).  Such objects were called by Peter Conti  (1984) \textit{Luminous Blue Variables} for the obvious reasons, which are actually not so obvious: these stars are often not Blue, and sometimes not Variable. LBVs are believed to be extremely massive stars evolving to the Wolf-Rayet stage.  The canonical Galactic examples are P Cygni and \ec;  a nice recent compendium of Galactic LBVs has been presented by \citet{clark}.  It's not completely clear how much mass is lost in these eposidic LBV eruptions compared to (relatively) steady stellar wind mass loss.  Both observational astronomers and theoreticians suspect not much, not because these eruptions aren't spectacular (\ec's for example released as much energy as a minor supernova) but because they don't seem to last very long.  But there have been interesting claims that for Population III objects perhaps such ejections are the dominant mode of mass loss, and if so such eruptions would have important implications on seeding the early Universe with heavy elements and black holes.  

Some questions are: how important are these LBV eruptions in determining the ultimate fate of a massive star?  and which massive stars undergo such eruptions? and how often do they occur (or recur)? and how do they depend on changes in angular momentum and magnetic fields? or do they help drive changes in angular momentum and magnetic fields? And what causes these eruptions anyway? 

Duplicity also undoubtedly plays a major role in the process of evolution, at least for those systems with close companions (which seem to be, if not the majority of massive stars, then a substantial fraction) and possibly even well-separated systems if the orbits are eccentric and periastrons close.  

\section{X-ray Emission: A Poor Probe of the LBV Phenomenon}
\label{sec:xray}

LBVs are surrounded by the detritus of their eruption.  This ejecta can be very thick, in many cases making direct observation of the LBV difficult.  Radiation which can penetrate the murk is useful as a probe of conditions inside.  Long-wavelength radiation is useful but limited by the extended size of the free-free photosphere (which can be a few AU in radius).  Hard X-radiation (above a few keV) can penetrate through enormous columns of material, in principle probing the innermost regions of the LBV wind. The difficulty is that you need a source of hard X-rays, and such sources are (unfortunately) hard to come by.  For example, P Cygni is an extremely weak X-ray source, as are most Galactic LBVs. We know of no LBV+ X-ray emitting collapsed companion system (and, in point of fact, very few Wolf-Rayet+collapsed systems either).  This makes X-ray studies a poor probe of LBVs in general.  However there are particular instances of  X-ray bright LBVs, and in these cases X-ray emission acts as a fine scalpel to dissect what's going in the hearts of these extreme stars. 

\subsection{\ec}

\ec\ is a well-known Galactic LBV; a relatively nearby (2300 pc) bright star which became enormously brighter in the 19th century in an event known in astronomical lore as the ``Great Eruption'', the residue of which can be seen as a structured, bipolar nebula (whimsically known as the ``Homunculus\footnote{``An artificially made dwarf, supposedly produced in a flask by an alchemist'' according to \url{dictionary.com}; in this case the Homunculus is actually the flask itself.}'') surrounding the star. Interferometry shows dense structured ejecta down to $0.1''$ ($\sim 200$ AU) or less from the star.  

\begin{figure}[!t]
  \includegraphics[width=\columnwidth]{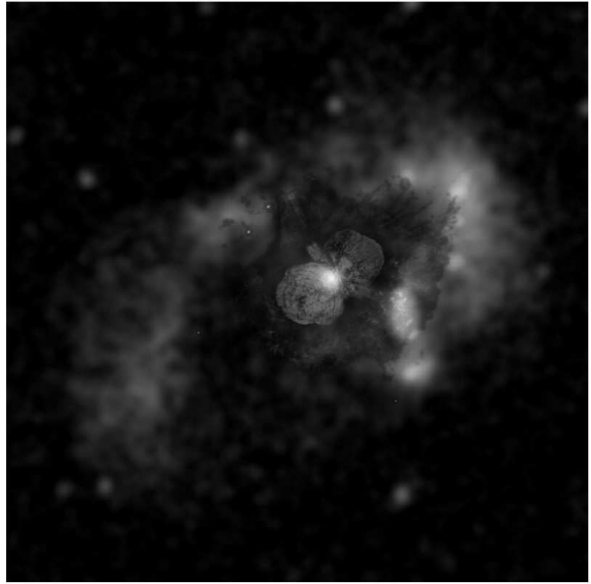}
  \caption{Comparison of a \chandra\ X-ray image of \ec\ and the Homunculus with an HST/WFPC2 image. The X-ray emission is extended and surrounds the inner bipolar Homunculus nebula.  \ec\ is the optical- and X-ray-bright point at the center of the image  WFPC2 image courtesy of N. Smith and J. Morse.}
  \label{fig:ecxo}
\end{figure}


\begin{figure*}[!t]
  \includegraphics[width=\textwidth]{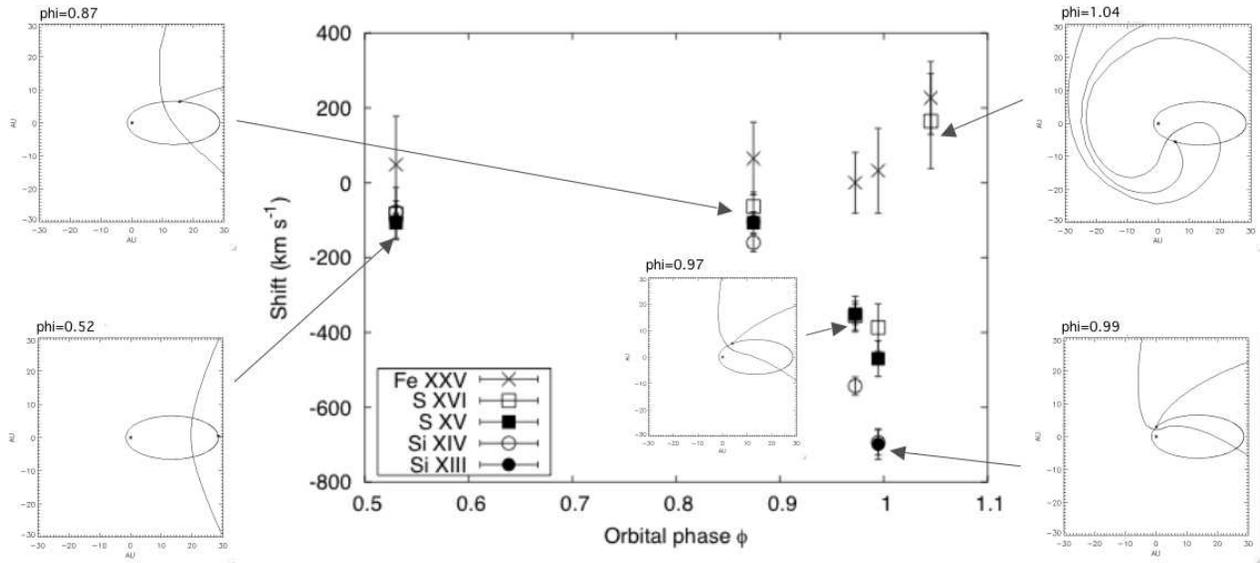}
  \caption{The graph shows X-ray emission line centroid velocities from \chandra\ High Energy Transmission Grating Spectrometer (HETGS) spectra as a function of X-ray cycle phase $\phi$.  Inset plots show the orientation of $\eta$~Car~B relative to $\eta$~Car~A and the shape of the contact discontinuity, with the observer to the right (in the direction of apastron, $\phi=0.5$) in all the plots.  The contact discontinuity shifts with the orbital motion, and is twisted by the Coriolis force near periastron when the orbital velocity of the secondary becomes comparable to the wind velocity of the primary. The highest X-ray line velocities observed occur when the trailing side of the shock cone (and the flow of the shocked gas along the cone) becomes more directed towards the observer.}
  \label{fig:ecxprof}
\end{figure*}

Periodic spectrometric variability and broad-band ($2-10$ keV) X-ray variations strongly suggest that the star is a colliding wind binary.  Not much is known about the companion, since it's difficult to detect directly.  Arguably the best implicit detection of the companion is through the X-ray emission\footnote{Note that \citet{rosina} claimed a spectral signature of the companion in far UV FUSE spectra, which might be an even better detection, though this result is still somewhat controversial.}  generated by the collision of the companion's wind with the wind of \ec\ (or more properly \ec~A).  This emission requires a wind velocity of $\sim 3000$ km s$^{-1}$ and a mass loss rate of $10^{-5}$ M$_{\odot}$ yr$^{-1}$, implying that the companion (\ec~B) is a bright supergiant or perhaps even a Wolf-Rayet star (in order to have a sufficiently fast, dense wind).   A guess as to the stellar parameters is given in Table \ref{tab:ecparams}.  These numbers are largely taken from \citet{hillier01}, \citet{corc01}, \citet{hotp},  \citet{katya05} and \citet{corc05}.  

The shocked thermal gas produces line emission from simple helium-like and hydrogen-like ions, offering numerous important diagnostics of the conditions of the shocked gas, both dynamic and thermodynamic. High resolution transmission grating spectra can be used to measure line centroids and thus the bulk flow of the shocked gas.  Centroids measured from strong lines (Si XIII \& Si XIV, and S XV \& S XIV in particular) are shown in Figure \ref{fig:ecxprof}, along with a simple model at each phase of the changing orientation and geometry of the ``contact discontinuity'', the boundary which separates the strong (slow) wind of \ec\ A from the weak (fast) wind of \ec\ B.  Significant variations in the line centroid velocities are dominated by the projected velocity of the flow along the line of sight when the orientation of the flow changes as the companion moves in orbit.

\begin{table}[!t]\centering
  \setlength{\tabnotewidth}{\columnwidth}
  \tablecols{4}
  \caption{Estimated System Parameters for \ec} \label{tab:ecparams}
  \begin{tabular}{lccc}
    \toprule
    Parameter & \multicolumn{1}{l}{$\eta$ Car A} & \multicolumn{1}{l}{$\eta$ Car B} & \multicolumn{1}{l}{System}\\
    \midrule
  Mass M$_{\odot}$ & $90$ & $30$??\\
   Radius R$_{\odot}$  & $150$ & $20$??\\
   Lumin. $10^{6}$ L$_{\odot}$ & $4$ & $0.9$?\\
   $T_{eff}$ kK   & $15$ & $34$?\\
  $\dot{M}$ M$_{\odot}$/yr & $10^{-4}-10^{-3}$ & $10^{-5}$?\\
$V_{\infty}$ km/s   & $500-1000$ & $3000$?\\
Period (d) &  & & $2024\pm2$\\
$e$ &  & & $0.8-0.95$\\
$a$ AU &  & & 15?\\
$i^{\circ}$ &  & & $45-90$\\
    \bottomrule
   \end{tabular}
\end{table}

\begin{figure}[!t]
  \includegraphics[width=\columnwidth]{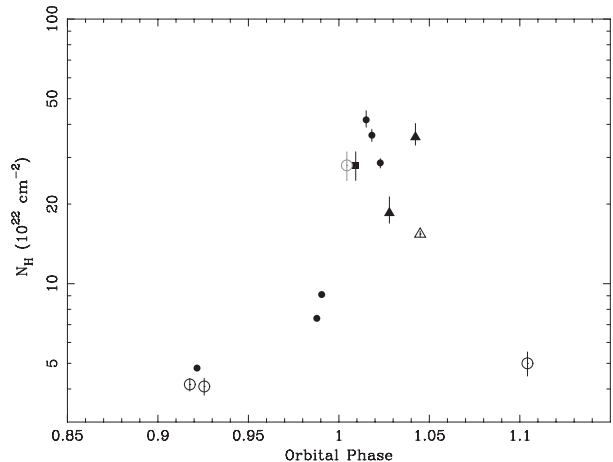}
  \caption{Column densities to the X-ray source vs. X-ray phase as measured by \xmm\ (filled circles), \chandra\ (filled triangles), \asca\ (open circles) and \sax\ (open triangles).  The column density reaches a maximum during the minimum, and again after the end of the minimum.  This might suggest either a pileup of material on the shock front, or perhaps a ``mini-ejection'' of material occurring near periastron passage. From \citet{kenji}.}
  \label{fig:ecnh}
\end{figure}

The variation of the X-ray spectrum through the X-ray low state has been discussed by \citet{kenji}. Among other results, they provide the first accurate measurement of the variation of the column density in front of the X-ray source during the minimum (see figure \ref{fig:ecnh}). The amount of material in front of the X-ray souce reaches a maximum during the X-ray brightness minimum, and again after the X-ray minimum ends.  This variation might suggest either a pileup of wind material from \ec\ A on the shock front, or perhaps even a ``mini-ejection'' of material occurring near periastron passage.

\subsection{HD 5980}

\begin{figure}[!t]
  \includegraphics[width=\columnwidth]{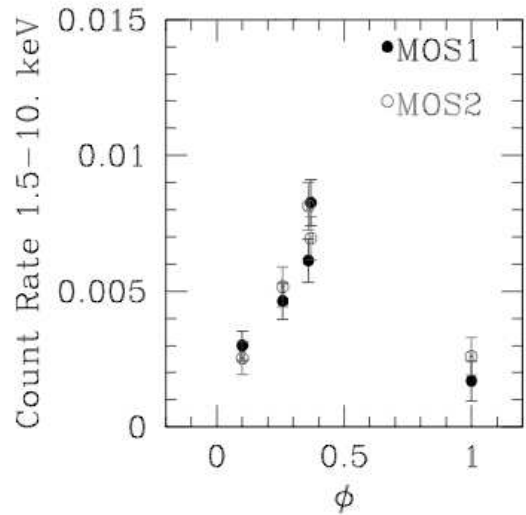}
  \caption{X-ray fluxes as a function of orbital phase for HD 5980, from \citet{yael}. The X-ray flux peaks during the eclipse of star A by star B, a WR star companion with a lower mass loss rate than star A.}
  \label{fig:hdlc}
\end{figure}

Briefly: HD 5980 is a massive, 20-day eclipsing binary in the Small Magellanic Cloud, one component of which (Star A) underwent an LBV-type eruption around 1994, while the other (Star B) is a WR star. The system is one of 12 known Wolf-Rayet stars (actually 13 depending on the state of Star A) in the SMC.   A good summary of the system is available from \citet{gloria}. \citet{yael} have recently shown that the X-ray emission from the system is variable and phase-locked to the orbit, the first time that phase-dependent X-ray emission has been seen in a massive binary beyond the Milky Way.  The phase-dependence seems rather strict despite a change in the LBV's mass loss rate by about a factor of 5 over the time interval of the X-ray observations.  Figure \ref{fig:hdlc} shows the X-ray lightcurve from \xmm\ observations in the $1.5-10$ keV band.  Interestingly, the X-ray brightness at phase $\phi=0.36$ (secondary eclipse) has remained the same in observations separated by about five years, even though the mass loss rate from Star A has declined by a large amount over that time. This suggests that the decline in mass loss is compensated by an increase in the wind speed from Star A.  

\citet{yael} also showed weak evidence that the X-ray hardness of the system peaks near secondary eclipse along with the X-ray brightness.  If confirmed, this means that the X-ray flux increase is not simply due to the presence of extra soft emission which might be expected (since, at secondary eclipse, we're viewing the shock through the lower density wind of star B). Rather this means that when star B is in front we're seeing extra \textit{hard} emission.  This may mean that the weaker wind of star B allows more of the hottest part of the shock to be viewed at this phase.  Because the system is eclipsing, this means that the hottest part of the shock cone must be larger than the photosphere of star B. An alternative is that, because the system is eccentric, it may be that the pre-shock wind velocities increase as the stars move towards apastron, resulting in harder emission. A test of this would be to view the X-ray emission of the system at apastron, an observation which has not yet been accomplished.

\section{Conclusions}

Massive stars are rare, and Luminous Blue Variables rarer still.  It is unclear how important this stage of evolution is, or exactly how massive a star needs to be to pass through it.  If sufficient mass loss is accomplished in this stage (either because the stage is sufficiently long, or mass loss sufficiently violent, or because stars may pass through the LBV phase multiple times before the supernova explosion) then it may play a major role in the evolution of massive stars.  It may be, as suggested by \citet{nathanstan}, that eruptive, LBV mass loss plays an especially important role at low metallicities where stellar wind driving is not so effective.  Such giant eruptions could play a significant role in affecting the evolution of Population III-type stars.
The low-primordial-metallicity SMC LBV HD 5980 may be an interesting test case of this proposition. 

\ec\ is probably the best studied LBV; but has all this study led us to a deeper understanding of the LBV phenomena, or is \ec\ a ``gonzo'' oddball?  Of course, in a class as small as the class of LBVs, it's hard to draw \textit{any} general conclusions.  The ``discovery'' of the companion star (if it can really be considered ``discovered'') perhaps points the way to deeper understanding of the LBV phenomena, if duplicity is fundamental to it (and it's been suspected that duplicity may play a role in shaping bipolar nebulae like the Homunculus).  On the other hand if duplicity is an ancillary trait of LBVs, then \ec~B provides a rare, in situ probe of the LBV experience.   

\bigskip 
\noindent In memory of Virpi, who touched so many lives:\\
\emph{---Si alguien ama a una flor de la que s\'olo existe m\'as que un ejemplar entre los millones y millones de estrellas, es bastante para que sea feliz cuando mira a las estrellas. Puede decir satisfecho: ``Mi flor est\'a all\'i, en alguna parte...''}\\
de Saint Exup\'ery, \textit{The Little Prince}


\bigskip 

\begin{thebibliography}

\bibitem[Clark et al.(2005)]{clark} Clark, J.~S., Larionov, 
V.~M., \& Arkharov, A.\ 2005, \aap, 435, 239 

\bibitem[Corcoran et al.(2001)]{corc01} Corcoran, M.~F., 
Ishibashi, K., Swank, J.~H., \& Petre, R.\ 2001, \apj, 547, 1034 

\bibitem[Corcoran(2005)]{corc05} Corcoran, M.~F.\ 2005, \aj, 
129, 2018

\bibitem[Hillier et al.(2001)]{hillier01} Hillier, D.~J., 
Davidson, K., Ishibashi, K., \& Gull, T.\ 2001, \apj, 553, 837 

\bibitem[Hamaguchi et al.(2007)]{kenji} Hamaguchi, K., et al. 2007, \apj, in press. \\
\url{http://arxiv.org/abs/astro-ph/0702409}

\bibitem[Iping et al.(2005)]{rosina} Iping, R.~C., Sonneborn, 
G., Gull, T.~R., Massa, D.~L., \& Hillier, D.~J.\ 2005, \apjl, 633, L37 

\bibitem[Koenigsberger(2004)]{gloria} Koenigsberger, G.\ 2004, 
Revista Mexicana de Astronomia y Astrofisica, 40, 107 

\bibitem[Naz\'e et al.(2007)]{yael} Naz\'e. Y., et al. 2007, \apjl, in press \\ \url{http://arxiv.org/abs/astro-ph/0702403}

\bibitem[Pittard \& Corcoran(2002)]{hotp} Pittard, J.~M., \& 
Corcoran, M.~F.\ 2002, \aap, 383, 636 

\bibitem[Smith \& Owocki(2006)]{nathanstan} Smith, N., and Owocki, S. P. 2006, \apjl, 645, 45.

\bibitem[Verner et al.(2005)]{katya05} Verner, E., Bruhweiler, 
F., \& Gull, T.\ 2005, \apj, 624, 973 

 
\end{thebibliography}
\end{document}